\documentclass{rs-arxiv}

\usepackage{epsfig,color}                     
\usepackage{graphicx}
\graphicspath{ {images/} {../plos/}  {../images/} {../plos/images/} }

\usepackage{graphicx}
\usepackage{amsmath}
\usepackage{amssymb}
\usepackage{amsfonts}
\usepackage{amsthm}
\usepackage{endfloat}
\usepackage[numbers]{natbib}
\usepackage{endnotes}
\usepackage{setspace}
\usepackage{verbatim}
\usepackage[left=1in, right=0.7in, bottom=0.4in, top=1in]{geometry}
\usepackage{times}
\usepackage{helvet}
\usepackage{courier}
\usepackage{bm}
\usepackage{url}
\usepackage{babel}
\usepackage{dcolumn}	
\usepackage{multirow}

\newcommand{\SIR}{$SIR$}
\newcommand{\SI}{$SI$}
\newcommand{\SIS}{$SIS$}
\newcommand{\SEIR}{$SEIR$}
\newcommand{\R}{$\mathcal{R}_0$}

\newcommand{\BDSIR}{BDSIR}

\begin{document}

\title{Simultaneous reconstruction of evolutionary history and epidemiological dynamics from viral sequences with the birth-death \SIR{} model}
\author{%
Denise K\"{u}hnert$^{1,2}\footnote{denise.kuehnert@env.ethz.ch}$, 
Tanja Stadler$^{3}$, 
Timothy G Vaughan$^{1,4}$, 
Alexei J Drummond$^{1,5}$}
\address{ 
$^1$Department of Computer Science,University of Auckland, Auckland, NZ;
$^2$Department of Environmental Systems Science, ETH Z\"{u}rich, Switzerland;
$^3$Department of Biosystems Science \& Engineering, ETH Z\"{u}rich, Basel, Switzerland;
$^4$Institute of Veterinary, Animal \& Biomedical Sciences, Massey University, Palmerston North, NZ and
$^5$Allan Wilson Centre for Molecular Ecology and Evolution, University of Auckland, Auckland, NZ
}

\abstract{
The evolution of RNA viruses such as HIV, Hepatitis C and Influenza virus occurs so rapidly that the viruses' genomes contain information on past ecological dynamics.  
Hence, we develop a phylodynamic method that enables the joint estimation of epidemiological parameters and phylogenetic history.
Based on a compartmental {susceptible-infected-removed} (\SIR{}) model, this method provides separate information on incidence and prevalence of infections. 
Detailed information on the interaction of host population dynamics and  evolutionary history can inform decisions on how to contain or entirely avoid disease outbreaks.

We apply our Birth-Death \SIR{} method (\BDSIR{}) to two viral data sets. 
First, five human immunodeficiency virus type 1 clusters sampled in the United Kingdom between 1999 and 2003 are analyzed.  
The estimated basic reproduction ratios range from 1.9 to 3.2 among the clusters.  
All clusters show a decline in the growth rate of the local epidemic in the middle or end of the 90's. 

The analysis of a hepatitis C virus (HCV) genotype 2c data set 
shows that the local epidemic in the C\'ordoban city Cruz del Eje originated around 1906 (median), coinciding with an immigration wave from Europe to central Argentina that dates from 1880--1920. 
The estimated time of epidemic peak is around 1970.
}


\maketitle

\section{Introduction}
The fast evolution of RNA viruses poses a challenge: their evolutionary processes are subject to ecological dynamics that occur on the same time scale 
\citep{Grenfell:2004cr, Kuhnert:2011vn}.  
Therefore a credible model of virus evolution has to take time-dependent ecological processes into account. 
In this work we present a method for Bayesian inference under a phylodynamic model that simultaneously estimates epidemiological parameters and reconstructs phylogenetic history.

Recent developments have provided us with extensive amounts of genomic data. In the case of human immunodeficiency virus (HIV), a number of countries, such as Switzerland \citep{Swiss-HIV-Cohort-Study:2010fk} and the United Kingdom \citep{Sabin:2004uq}, have sampled a large fraction of HIV-infected residents. 
Analysis of such data sets requires careful validation of methods.  
For example, standard coalescent models require {the population size to be constant or to vary deterministically. 
To accommodate stochastic population size changes} within phylogenetic reconstruction, a tree prior based on the birth-death process \citep{feller:1939,Kendall:1948} has been developed by \cite{Stadler:2010fj}.  

An extension of Stadler's birth-death-sampling model, the birth-death skyline plot (BDSKY) \citep{Stadler:2013fk}, allows for serially sampled data and rate changes over time. These rate changes through time  may reflect environmental changes such as new treatment strategies or behavior changes at different points in time.

{Host population dynamics can strongly affect viral transmission and evolution \citep{Grenfell:2004cr}.}
Therefore, modeling the underlying host population through compartmental models not only provides additional information on the viral outbreak, but also informs the estimates for evolutionary reconstruction. 
We show here that the birth-death skyline plot {can be parametrized to enable} the underlying population dynamics to be modeled as a compartmental susceptible-infected-removed (\SIR{}) model, a classic epidemiological model which accounts for changing host population composition \citep{kermack1927}. 

In the {Birth-Death \SIR{} method (\BDSIR{})} model presented in this paper we assume that a gene genealogy, i.e.~the phylogeny connecting the sampled sequences, represents the past  transmission history of the hosts (note that of course this transmission history is incomplete as many infected hosts may not be sampled). 
That is, an infected host corresponds to {a portion of} a single lineage in the phylogeny, {and, of the two child branches produced at a branching node, one represents the continuation of the donor infection whereas the other represents the new recipient}.

We introduce the \BDSIR{} model for estimating epidemiological parameters such as the basic reproductive number based on sequence data. The model approximates a classic {stochastic}  
\SIR{} model. In summary, our method works as follows. Trajectories of the number of susceptible, infected and removed individuals are provided by the \SIR{} model. Based on the trajectory of infected individuals, the average transmission rate in short time intervals throughout the epidemic is determined. The likelihood of the proposed sampled tree connecting the sequence data is then obtained based on these piecewise constant transmission rates using the birth-death-skyline model. This \BDSIR{} model is implemented into the Bayesian software framework BEAST2 (http://beast2.cs.auckland.ac.nz).

We then perform a simulation study showing the accuracy of the \BDSIR{} model. 
Applied to HIV-1 type B sequences sampled in the United Kingdom (UK), the method gives insight into the epidemic features of five local epidemics. 
Although it is common to model the infection dynamics of HIV with non-recovery ($SI$) models, here we model it as an \SIR{} model.
In countries like the United Kingdom, behaviour changes and commencement of treatment are expected to coincide with the sampling of HIV-positive individuals, which {can  imply}  the removal of the individual from the infectious pool \citep{Montaner:2006fk}.
{Finally, we apply the method to a set of HCV type 2c sequences from the city of Cruz del Eje in the Argentinian province C\'ordoba. 
European immigrations likely caused the outbreak of this local epidemic. Many of the immigrants came from Italy, where HCV subtype 2c is also common \citep{Re:2011uq}.
The epidemic appears to have peaked around 1970 and to be in its decline now.
}

\section{Methods} 

\subsection*{Stochastic epidemiological models}

Infectious disease epidemics are classically modeled through compartmentalization into a number of host compartments, such as susceptible, infected and removed individuals (\SIR{} model), where a susceptible individual moves to the infected compartment upon infection, and an infected individual moves to the removed compartment upon removal/recovery. 
Such a model may be extended by assuming an exposed class (\SEIR{} model), or altered by assuming no removal/recovery (\SI{} model) or no immunity of recovered individuals (\SIS{} model) \citep{Kermack:1932fk}.

In the following, we formalize a stochastic epidemiological \SIR{} model which we will use for phylogenetic inference {assuming an unstructured population}. 
It is relatively straightforward for other {unstructured} compartmental epidemiological models to be  placed into a stochastic framework for phylogenetic analysis in the same way.
 
In terms of its reaction kinetics a stochastic \SIR{} model has the following scheme:
\begin{eqnarray}
I + S & \overset{\beta} \longrightarrow & 2I \nonumber \\
I & \overset{\gamma}\longrightarrow & R 
\label{scheme1}
\end{eqnarray}
An individual in the infected compartment $I$ infects a susceptible individual $S$ at a mass-action infection rate of $\beta$. 
An infected individual $I$ recovers at recovery rate $\gamma$. 

Typically such \SIR{} models are formalized through a system of ordinary differential equations, which represent a mean field approximation of the expected number of susceptible, infected and removed {individuals} through time of a stochastic model,
with $n_S(0)$ susceptible individuals, {$n_I(0)$=}1 infected, and {$n_R(0)$=}0 removed individuals {as initial conditions} at time 0:

\begin{eqnarray*}
\frac{d}{dt} n_S(t) &=& - \beta n_S(t)  n_I(t) \\
\frac{d}{dt} n_I(t) &=& \beta  n_S(t) n_I(t) - \gamma n_I(t) \\
\frac{d}{dt} n_R(t) &=& \gamma n_I(t)
\end{eqnarray*}

Stochasticity plays a significant role in viral epidemics, especially at the very beginning of an epidemic.  Although large epidemics can be described by deterministic models once they are established, these deterministic models must condition on the time at which the exponential growth phase of the epidemic begins, since this starting time impacts the timing of every event thereafter.

Hence, we employ stochastic epidemiological models here. 
Under the stochastic \SIR{} model, an infected individual infects a susceptible individual with rate $\beta$ and recovers with rate $\gamma$.

In most epidemics, we only observe a proportion $s$ of the recoveries. 
We can include this by adding another reaction to Equation \ref{scheme1}:
\begin{eqnarray}
\label{sirSims}
I + S & \overset{\beta} \longrightarrow & 2I \nonumber \\
I & \overset{(1-s)\gamma}\longrightarrow & R_h \\
I & \overset{s\gamma}\longrightarrow & R_s, \nonumber
\end{eqnarray}

where we distinguish between {\it hidden} or unobserved recoveries $R_h$ and {\it sampled} or observed recoveries $R_s$. 
The  {\it sampling proportion} $s$ with $0 \leq s \leq 1$ is the probability of a recovery being observed, and thus the expected proportion of recoveries observed.
This infection process, where only some recoveries are observed, is the basis for connecting non-linear epidemiological models to phylogenetic data. 

Molecular sequence data from infected hosts, which is used in order to infer the phylogenetic tree, is often sampled sequentially through time. 
In our  model we account directly for this sequential sampling as an infected individual is sampled with rate $\psi =  s \gamma$, and upon sampling the individual moves to the removed class (due to e.g. successful treatment or behavior change).

\begin{figure}
\begin{center}
\includegraphics[width=8.3cm]{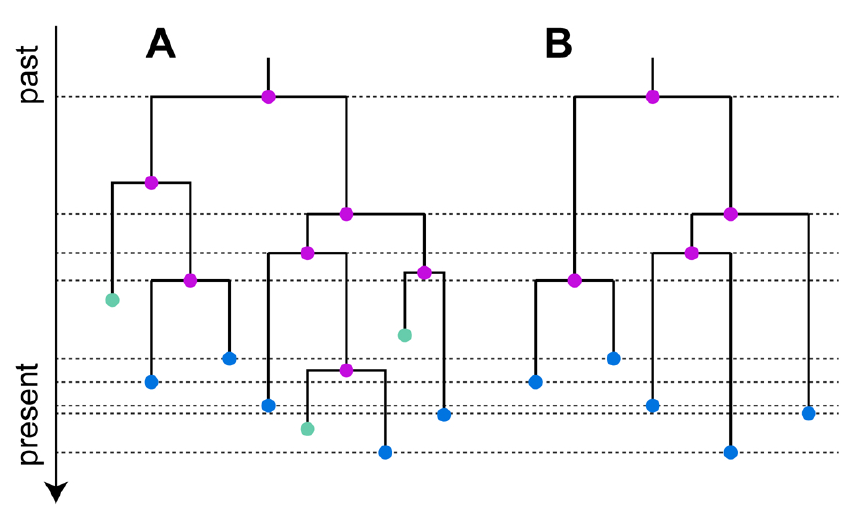}
\end{center}
\caption{
{\bf Sequentially sampled birth-death-sampling tree.}  (a) Full transmission tree with birth (purple), death (green) and sampling (blue) events. (b) Full tree pruned to only include observed, i.e.~sampled individuals.
}
\label{fig:bdstree}
\end{figure}

The stochastic \SIR{} model with transmission rate $\beta$, recovery rate $\gamma$, sampling proportion $s$, population size $n_S(0)$
and timespan of the epidemic being $T$  induces a distribution of full transmission chains through time (i.e. who infected whom).
The sampled tree (or sampled transmission chain) results from the full transmission chain by pruning all non-sampled lineages, i.e.~the tips of the sampled tree are the sampled individuals{, see Figure \ref{fig:bdstree}}.
The trajectories of the \SIR{} model are the time series of number of susceptible individuals, infected individuals and removed individuals through time.

{Note that we assume the host population size $N=n_S(i)+n_I(i)+n_R(i)$ to be constant over time, in which case our population-dependent model (transmission term $\beta n_S n_I$) is equivalent to a frequency-dependent model (transmission term $\frac{\beta}{N} n_S n_I$).}

\subsection*{Incorporating stochastic epidemiological models into phylogenetics}
\begin{figure}
\begin{center}
\includegraphics[width=8.3cm]{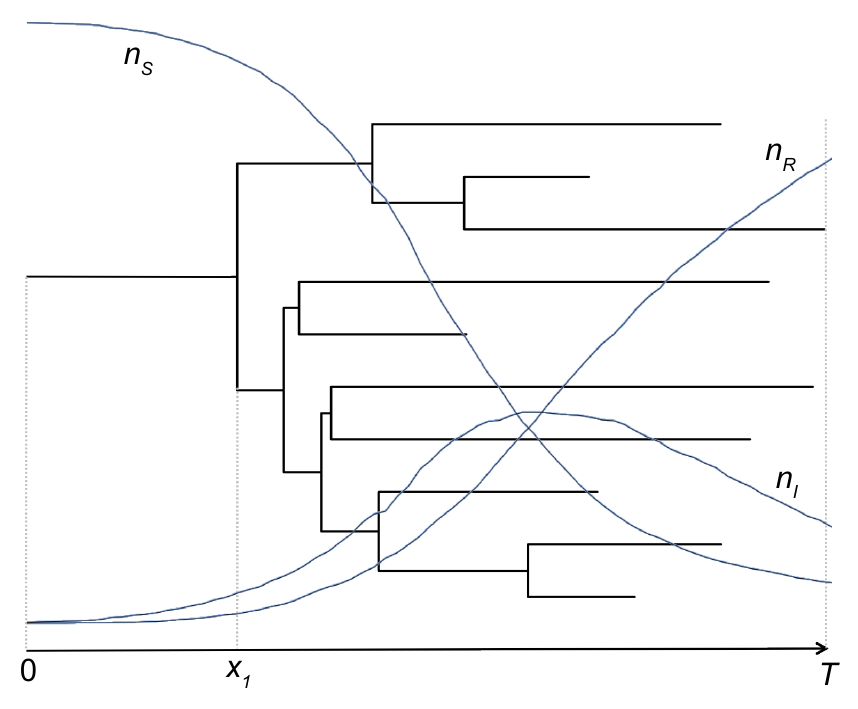}
\end{center}
\caption{
{\bf An epidemic starts at time $0$, giving rise to the genealogy rooted at time $x_1$, and trajectories for the number of susceptible ($n_S$), infected ($n_I$), and removed ($n_R$) individuals}.  The last sampled tip determines the end of the observed epidemic at time $T$.
}
\label{treesir}
\end{figure}

We do not have information about unobserved individuals, i.e. we cannot expect to infer the full transmission chain.
However, based on sequenced data $D$ from a sample of infected individuals,  we aim at inferring the sampled {transmission} tree $\mathcal{T}$, {the evolutionary parameters $\theta$,} 
the \SIR{} trajectories $${\mathcal{Y} = \{ Y_t = \{n_S(t), n_I(t), n_R(t) \}, 0 \leq t \leq T\}}$$ (where $Y_0 = \{n_S(0), 1,0 \}${, i.e.~initially all individuals are susceptible, apart from 1 individual, which is infected}) and the epidemiological parameters {$\eta=(\lambda,\mu,\psi,n_S(0),T)$, where $\lambda = \beta n_S(0)$, $\mu = (1-s) \gamma$ and $\psi =  s \gamma$}, in a Bayesian framework, see Figure \ref{treesir}. In particular, we want to infer the posterior distribution of trees,  trajectories, and parameters,
$$f(\mathcal{T}, \theta,\mathcal{Y},\eta | D) = \mathbb{P}(D|\mathcal{T}, \theta) f(\mathcal{T},\mathcal{Y}|\eta) f(\theta) f(\eta),$$
with $ \mathbb{P}(D|\mathcal{T}, \theta)$ being the likelihood of the sequences given a tree (which can be calculated efficiently with Felsenstein's pruning algorithm \citep{felsenstein:2004}) and $f(\theta)$, $f(\eta)$ being the prior distributions on the parameters.
Furthermore the inference requires the expression for the joint probability of the sampled tree and the trajectories given  the epidemiological parameters, $f(\mathcal{T},\mathcal{Y}|\eta)  $. 
We rewrite,
$$f(\mathcal{T}, \mathcal{Y} | \eta) = f(\mathcal{T} | \mathcal{Y} ,\eta)  f(\mathcal{Y} | \eta) .$$

The right hand side of the equation is the probability density of a sampled transmission tree given the trajectories and the epidemiological parameters, multiplied by the probability density of the trajectories given the epidemiological parameters.
Both terms must be determined so that we can do Bayesian phylogenetic inference under the stochastic \SIR{} model.

Instead of calculating $f(\mathcal{Y} | \eta) $, we can simulate a trajectory given  the epidemiological parameters  $\eta$ in each MCMC step (for details see Supporting Information). Given the simulated trajectory, it remains to calculate $f(\mathcal{T} | \mathcal{Y} , \eta)$. 

 For calculating the probability density of a sampled tree, we note that {when conditioning} 
 on the full trajectories, we have 
 $f(\mathcal{T} | \mathcal{Y} , \eta) = f(\mathcal{T} | \mathcal{Y} ),$
and the probability of a sampled tree given the trajectories, $f(\mathcal{T} | \mathcal{Y} )$, is  a product where at each event in the trajectories we multiply by the probability of {the event} having happened in the sampled tree  if it coincided with a tree event, and multiply by the probability of {the event} having not happened in the sampled tree if it did not coincide with a tree event. Thus theoretically we can both simulate trajectories and evaluate the tree probability $f(\mathcal{T} | \mathcal{Y} )$. 
For large population sizes (i.e. large $n_S(0)$), the number of events will grow very large, thus both trajectory simulations and tree likelihood calculation will become very slow.
{Therefore, we do not substitute $f(\mathcal{T} | \mathcal{Y} , \eta) $ by $f(\mathcal{T} | \mathcal{Y} )$.}
{Instead, w}e approximate both the simulation and the likelihood calculation by discretizing time. 
With the simulation techniques described in Text S2 we simulate at discrete time points $t_1,t_2,\ldots,t_m$ where $t_i = \frac{iT}{m}$, the number of susceptible, infected and removed individuals, i.e.~we have trajectories \linebreak
$\mathcal{\tilde{Y}} = \{ \{n_S(0),n_I(0),n_R(0)\}, \ldots, \{n_S(m),n_I(m),n_R(m)\}\}$, with the initial value at time $0$ being $\{n_S(0),1,0\}$.
Then, we need to calculate
$f(\mathcal{T} | \mathcal{\tilde{Y}}, \eta).$

We note that so far, for $m \rightarrow \infty$, convergence to the exact probability densities holds.
However, we did not find an efficient way to calculate the required probability density $f(\mathcal{T} | \mathcal{\tilde{Y}}, \eta)$, thus we introduce an approximation below, yielding the \BDSIR{} model, which does not converge to the exact probability density, but turns out to be efficient and accurate.

We sample trajectories $\mathcal{\tilde{Y}}$ from $f(\mathcal{\tilde{Y}} | \eta) $ with a $\tau$-leaping algorithm (see Text S2).

\subsection*{The \BDSIR{} model}
The \BDSIR{} model is an approximate stochastic epidemiological model in phylogenetics.  
We approximate the stochastic \SIR{} model by the \BDSIR{} model, leading to an efficient way for calculating approximatively the likelihood of the phylogeny given the epidemiological  time series and parameters $f(\mathcal{T} | \mathcal{\tilde{Y}} , \eta)$.

In the \BDSIR{} model, the epidemiological trajectories are defined stochastically by the \SIR{} model {with constant population size ($n_S(i)+n_I(i)+n_R(i)$)}, and simulated using the $\tau$-leaping approach described in Text S2.
Simulations are started with an initial number of susceptibles, $n_S(0)$, and last for time $T$. 
At equally spaced time points $t_1,\ldots, t_m$ the values of the trajectories $n_S(i),n_I(i),n_R(i)$ are recorded, yielding $f(\mathcal{\tilde{Y}} | {\lambda,\mu, \psi},n_S(0),T) $. The trajectories converge to \SIR{} trajectories $\mathcal{Y}$ for $m \rightarrow \infty$. 

Under the \BDSIR{} model, a sampled tree is  induced by a so-called birth-death skyline plot \citep{Stadler:2013fk} given the discrete time trajectories $\mathcal{\tilde{Y}}$ as follows. 
The transmission rate $\lambda_i$ during time interval $[t_i,t_{i+1})$ is parametrized by $\lambda_i = \beta n_S(i)$, where $\beta$ is the epidemiological transmission rate and $n_S(i)$ is the number of susceptibles at time $t_i$. 
The recovery rate $\gamma$ and
 sampling fraction $s$ are constant through time. 
Piecewise constant transmission rates in the \BDSIR{} model allow the calculation of the likelihood of a sampled tree {$\hat{f}(\mathcal{T} | \{\lambda_i=\beta n_S(i) | i=0\dots m\},{\mu,\psi},n_S(0),T)$}.
{This likelihood} is given by the probability density of the birth-death skyline plot {(for a derivation of the probability density see \citep[Theorem 1]{Stadler:2013fk})} {with piecewise constant transmission rate $\lambda_i = \beta n_S(i)$ and constant death and sampling rate $\mu$ and $\psi$, respectively.} { The equation for the probability density of a sampled tree is stated in Text S1, Supporting Information}.

{In} the \BDSIR{} model, we approximate the calculation of the posterior distribution under the stochastic \SIR{} model, 
\begin{equation}
\label{posterior}
f ( \mathcal{T}, \mathcal{\tilde{Y}}, \eta | D) \propto \mathbb{P}(D | \mathcal{T}) f( \mathcal{T} | \mathcal{\tilde{Y}}, \eta) f(\mathcal{\tilde{Y}} | \eta) f(\eta)
\end{equation}

by using 

\begin{equation}
\label{approximation}
 f(\mathcal{T} | \mathcal{\tilde{Y}}, \eta)  \approx  \hat{f}(\mathcal{T} | \{\lambda_i=\beta n_S(i) | i=0\dots m\},{\mu,\psi},n_S(0),T).
\end{equation}

While $f(\mathcal{T} | \mathcal{\tilde{Y}}, \eta)$ converges to $ f( \mathcal{T} | \mathcal{\tilde{Y}})$ as $m \rightarrow \infty$, the approximation (Equation \ref{approximation}, RHS) does not: under the skyline plot, we only specify the transmission rates based on $\mathcal{\tilde{Y}}$. Based on these time-varying transmission rates, we calculate the likelihood of the tree by integrating over all possible trajectories $\mathcal{Y}$ yielding the given tree (instead of conditioning on $\mathcal{\tilde{Y}}$).

\subsection*{MCMC implementation of the \BDSIR{} model}
We implemented Equations \ref{posterior} \&  \ref{approximation} into BEAST for joint phylogenetic tree and epidemiological parameter inference (code and examples can be downloaded from http://code.google.com/p/phylodynamics).
The prior distribution $f(\mathcal{\tilde{Y}} | \eta)$ in Equation \ref{posterior} is subsumed in the proposal kernel of a Markov chain Monte Carlo (MCMC) implementation, so that a new trajectory $\mathcal{\tilde{Y}^\prime}$ is proposed by simulation, whenever a new $\eta^\prime$ is proposed giving a joint proposal kernel of:
\begin{equation}
q(\eta^{\prime}, \mathcal{\tilde{Y}}^{\prime} | \eta,  \mathcal{\tilde{Y}}) = q(\eta^\prime | \eta)f(\mathcal{\tilde{Y}^\prime}| \eta^\prime).
\nonumber
\end{equation}

Therefore, \BDSIR{} utilizes an independence Metropolis-Hastings (MH) sampler, as introduced by \cite{Stephens:2001qy} and subsequently studied by many others, e.g.~\citep{Beaumont:2003yq,Andrieu:2009ys}.
This leads to the Metropolis-Hastings acceptance ratio \citep{Hastings:1970zr}
{\scriptsize
\begin{eqnarray}
\alpha &=& \min\left(1,
\frac
{ \mathbb{P}(D | \mathcal{T}^{\prime}) f( \mathcal{T}^{\prime} | \mathcal{\tilde{Y}}^{\prime}, \eta^{\prime}) f(\mathcal{\tilde{Y}}^{\prime} | \eta^{\prime}) f(\eta^{\prime})}
{ \mathbb{P}(D | \mathcal{T}) f( \mathcal{T} | \mathcal{\tilde{Y}}, \eta) f(\mathcal{\tilde{Y}} | \eta) f(\eta)}
\times
\frac{q(\eta, \mathcal{\tilde{Y}} | \eta^{\prime},  \mathcal{\tilde{Y}}^{\prime}) }{q(\eta^{\prime}, \mathcal{\tilde{Y}}^{\prime} | \eta,  \mathcal{\tilde{Y}}) }
\right) 
\nonumber\\
 &=& \min\left(1, \frac
{ \mathbb{P}(D | \mathcal{T}^{\prime}) f( \mathcal{T}^{\prime} | \mathcal{\tilde{Y}}^{\prime}, \eta^{\prime})  f(\eta^{\prime})}
{ \mathbb{P}(D | \mathcal{T}) f( \mathcal{T} | \mathcal{\tilde{Y}}, \eta)  f(\eta)}
\times
\frac{q(\eta | \eta^{\prime})}{q(\eta^{\prime} | \eta)}
\right),
\nonumber
\end{eqnarray}
}
where $\eta^{\prime}$ denotes the new proposal of parameters $\eta$, etc.
The factor $f(\mathcal{\tilde{Y}}| \eta)$ is implicitly included in the posterior through independence sampling of the time series $\mathcal{\tilde{Y}}$.
The proposal is rejected, if at any of the times $t_i$, $i=0 \dots m$, the number of infected individuals in the proposed trajectory is less than the corresponding number of lineages in the phylogenetic tree.

In our simulation study we show that the approximation for the tree likelihood (Equation (\ref{approximation})) is suitable, by illustrating that we can infer parameters from simulated phylogenies with high accuracy.
Thus, applying our \BDSIR{} model to virus sequence data from different infected individuals throughout an epidemic, the phylogenetic tree can be estimated jointly with the epidemiological parameters $\eta$.
The choice of Bayesian parameter prior distributions is facilitated by the parameterization of the epidemiological parameters as the basic reproduction ratio $\mathcal{R}_0 = \frac{n_S(0) \beta}{\gamma}$, the rate at which infected individuals become noninfectious $\gamma$, the sampling proportion $s$, the initial susceptible population size $n_S(0)$ and the length of the epidemic $T$.

\subsection*{Simulation study}

Using simulations we explore how well the \BDSIR{} model performs when inferring parameters based on simulated trees. 
In {Stage 1} we simulate $100$ \SIR{} trees based on the reaction scheme (\ref{sirSims}) with {$n_S(0)=999$,} $\beta= 0.00075$, $\gamma=0.30$ and $s=\frac{1}{6}$ (i.e.~$\mathcal{R}_0 = 2.5$). 
Each simulated tree has $100$ tips.
Then, we set up an analysis to re-estimate the simulation parameters for each of the simulated trees.
In this second stage, the tree and the duration $T$ of the epidemic are fixed; they represent the data from which we estimate the epidemiological parameters. 

Stage 2 is comprised of two $\times$ two sets of analyses: In the first two sets, we fixed the sampling proportion $s$ as we showed in \citep{Stadler:2013fk} that $\lambda$, $\gamma$ and $s$ correlate; in the second two sets, we estimated $s$.
In each set of two, the initial number of susceptible individuals $n_S(0)$ is firstly fixed to the true value and secondly all parameters including $n_S(0)$ are estimated. 
We chose $m=100$ {equidistant time points $t_1, t_2, \dots , t_m$ to discretize} the epidemic trajectories.
For comparison, we also estimate the rates of the second two sets (i.e. estimating $s$) with (i) the birth-death skyline model \citep{Stadler:2013fk} with piecewise constant effective reproduction ratio and (ii) the birth-death-sampling model with constant effective reproduction ratio \citep{stadler:2012}. 

While the birth-death-sampling model characterizes the tree-generating process through constant birth, death and sampling rates, these rates can change in a piecewise fashion in the birth-death-skyline model. 
Both methods differ from the \BDSIR{} model in that they do not explicitly parametrize the underlying host population dynamics. 
We compare the estimated parameters to the true parameter values. In particular, we focus on the {\em basic reproduction ratio} \R{} (the average number of secondary infections in a completely susceptible population), and the {\em effective reproduction ratio} (the average number of secondary infections in the current population).

The \BDSIR{} method estimates the basic reproduction ratio as $\mathcal{R}_0 = \frac{\beta n_S(0)}{\gamma}$.
BDSKY estimates the effective reproduction ratio $\mathcal{R}_i$ for each time interval $[t^{\prime}_i,t^{\prime}_{i+1})$.
We chose $10$ intervals for the BDSKY analysis, such that  $t^{\prime}_i = \frac{iT}{10}$. 
We obtained the `true' effective reproduction ratio from the Stage 1 simulations {of the \SIR{} trees} (as well as the estimates for \BDSIR{}) by 
 computing the averaged effective reproduction ratios $\overline{\mathcal{R}}_i = \frac{\beta \cdot \overline{n_S(i)} }{\gamma}$, $i=1..10$, (where $\overline{n_S(i)}$ is the mean number of susceptible individuals, given by true trajectory $\mathcal{\tilde{Y}}$ in time interval $[t^{\prime}_i,t^{\prime}_{i+1})$).

{Relative error, bias and highest posterior density (HPD) width served as measures of precision and accuracy. 
We define the relative error as 
$$error =\frac{|\overset{\wedge}{\eta}_{median} - \eta|}{\eta}, $$
the relative bias as
$$bias =  \frac{\overset{\wedge}{\eta}_{median} - \eta}{\eta}, $$
and, finally, the $95\%$ relative HPD width is defined as 
$$\frac{95\% \text{ HPD upper bound} - 95\% \text{ HPD lower bound}}{\eta}, $$
where $\eta$ is the true parameter and $\overset{\wedge}{\eta}_{median}$ is the posterior median value of the parameter.}

The Bayesian prior distributions used in Stage 2 are given in Table \ref{table:priors}.

\subsection*{HIV-1 type B in the UK}
A set of molecular sequences sampled from HIV-1 type B infected individuals in the UK, have been grouped into five phylogenetic clusters \citep{Hue:2005fk}. 
Sampled between 1999-2003 these clusters represent a suitable example data set for the analysis under the \BDSIR{} model. 
The clusters are comprised of 41, 62, 29, 26 and 35 sequences, respectively, and correspond to clusters 1--4 and 6 in the original analysis. 
{{Each cluster is considered as a sample from a local sub epidemic. 
Our model explicitly accounts for incomplete sampling of the local epidemics.}
These clusters have been identified based on a phylogenetic neighbor-joining tree that was constructed from 3,429 HIV-1 subtype B pol gene sequences from the UK and throughout the world. 
Note that the clusters are therefore not randomly sampled, and we also cannot guarantee that the sample sets are truly isolated transmission clusters. Although this identification of transmission clusters is common practice, we point out that it may introduce a bias.

}

Note that we employ an \SIR{} model, although true recovery in the literal sense does not (yet) occur in HIV-infected individuals. 
This is reasonable in countries like the United Kingdom, due to changes in behaviour as well as due to the effects of combination drug therapy, which can reduce viral load to undetectable levels, severely diminishing risk of further transmissions and, hence, implying removal of the individual from the infectious pool.
{However, during the earlier part of the study period, i.e.~before the introduction of HAART, this does not hold. }
Furthermore, modeling the HIV host population dynamics as a {closed} \SIR{} compartmental model requires to assume that the times at which individuals move between compartments are exponentially distributed {and that the host population size remains constant over time}. Another implicit simplifying assumption is that infected individuals are constantly infectious.

The phylodynamic analysis employed a {general time reversible substitution model with gamma distributed rate heterogeneity and a proportion of invariant sites (GTR+$\Gamma$+I)}, and all parameters were estimated jointly, apart from the substitution rate, which was fixed to $2.55 \times 10^{-3}$ as in \citep{Hue:2005fk}.
Before 1999 we assume the sampling proportion $s$ to be zero, since all samples were collected between 1999 and 2003. 

\subsection*{HCV type 2c in Argentina}
{
We analyze a set of 44 HCV type 2c sequences (NS5B region) that were sampled in 2004 during a survey in the city of Cruz del Eje (CdE), in C\'ordoba province, Argentina.
According to the survey, the 44 sequences included here represent roughly $2.8\%$ of the HCV-2c infected individuals in CdE, which has a population size of about 35000 and a proportion of $90\%$ genotype 2c infections out of all HCV positive patients encountered during the survey \citep{Mengarelli:2006yq}. 
Genotype 2c was probably introduced to Argentina during a European immigration wave between 1880--1920 \citep{Re:2011uq}.
A superset of this data (with additional samples from C\'ordoba province) was recently analyzed by \cite{Dearlove:2013ys} and in their model comparison they found that the SIR model is most suitable for this data.
The analysis employed a GTR+$\Gamma$+I substitution model and a strict clock model with the substitution rate fixed to  $0.58 \times 10^{-3}$ \citep{Tanaka:2002fk}.
Since all sequences were sampled at one time point (i.e.~homochronously), we model the sampling process through a sampling probability $\rho$ \citep{Stadler:2013fk}. 
This means that at the end of the tree (e.g.~in 2004) each infected individual was sampled with probability $\rho$.
}

In all analyses \SIR{} trajectories were sampled on $m=100$ intervals.
Table \ref{table:priors} shows the choice of Bayesian prior distributions for the analyses.

\section*{Results}
\subsection*{Simulation study}
\begin{figure} 
\begin{center}
\includegraphics[width=0.9\linewidth]{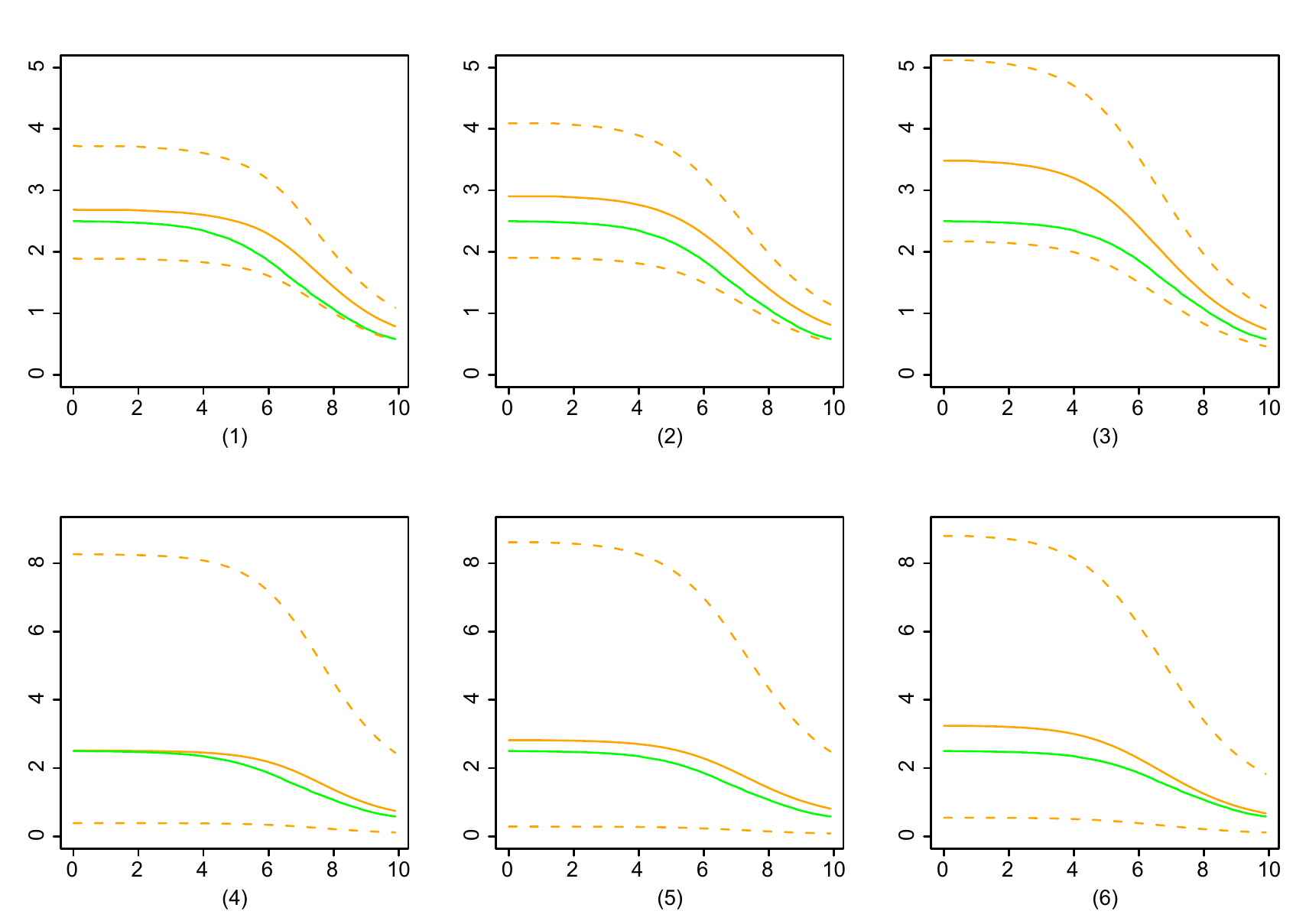} 
\caption{{\bf Reconstructed effective reproduction ratio from simulated \SIR{} trees.} True trajectory (green) versus estimated trajectory (orange) with 95\% HPD (dashed lines). Random sample of the 100 reconstruction results shown with $n_S(0)$ fixed to the true value (1-3) and estimated (4-6). Estimation of $n_S(0)$ throughout the phylodynamic reconstruction results in broader HPD intervals. }
\label{fig:bdsir_sim_trajs}
\end{center}
\end{figure}

We investigated the accuracy of our method through a simulation study. Based on reaction scheme (\ref{sirSims}) 100 serially sampled trees were simulated and then used for re-estimation of the simulation parameters. All four sets of analyses, (1) \BDSIR{} with fixed $n_S(0)$, (2) \BDSIR{}, (3) birth-death skyline (BDSKY) with $m=10$ intervals (i.e.~9 rate changes) and (4) birth-death-sampling, resulted in accurate estimates of the corresponding simulation parameters, or their time-averages (Tables \ref{table:SIR_BDSIR_fixS} - \ref{table:SIR_BDSKY_1}). 
Figure \ref{fig:bdsir_sim_trajs} shows trajectories of the reconstructed reproduction ratio for three simulations (randomly chosen from the set of 100 simulations). 
As one would expect, estimating the initial number of susceptible individuals $n_S(0)$ rather than fixing it to the true value results in broader 95\% HPD intervals. 

The epidemic dynamics were recovered well for {all} three analysis sets $(1)-(3)$.
{A slight positive bias in the estimates of the reproduction ratios is observed, which we speculate is due to the approximation employed by this method. This bias is small for low reproduction ratios $(R_0 < 5)$, where demographic stochastic effects are relevant, and the coverage properties of the estimator show that the uncertainty in the estimates is accurate.  This bias increases with higher $R_0$ (data not shown), suggesting that the \BDSIR{} method is most appropriate for modelling epidemics with low to moderate reproduction ratios ($R_0 < 10$). }
The effective reproduction ratio $\overline{\mathcal{R}}_1$ near the origin of the epidemic is estimated with the smallest bias among all $\overline{\mathcal{R}}_i$, $i=1 \dots 10$, respectively. 
Analysis under BDSKY results in the broadest relative HPD for $\overline{\mathcal{R}}_1$. 
Moving towards the present, the HPD interval widths for BDSKY mainly decrease.
The uncertainty in the epidemic dynamics suppresses this effect in the \BDSIR{} analyses: the relative HPD widths of the computed averages $\mathcal{\overline{R}}_i$ vary only slightly among the time intervals. 
Overall, the \BDSIR{} analyses with $n_S(0)$ fixed to the true value obtains the narrowest HPD intervals, yet, error rates and HPD accuracy are best, when $n_S(0)$ is estimated.

The birth-death-sampling model, which is equivalent to a one-dimensional BDSKY model, estimates the time averaged reproduction ratio accurately with quite narrow HPD intervals, suggesting it may be a reasonable method for inference in scenarios where the epidemic dynamics over time are not important. 

As shown by \cite{Stadler:2013fk} the parameters \R{}, $\gamma$ and $s$ of a birth-death-sampling tree prior are correlated. Therefore, we performed an additional set of simulations in which the sampling proportion $s$ is fixed to the true value. As expected, this results in narrower HPD intervals with accurate estimates of \R{} and $\gamma$. The HPD for the initial number of susceptible individuals $n_S(0)$ contains the true value, but is fairly wide as before (Supporting Information, Tables S1-S4).
These simulation results suggest that additional information about the pathogen under investigation can improve the parameter estimates of the \BDSIR{} analysis. In the case of HIV, for example, many countries have good estimates of how much of the infected population has been sampled.

\subsection*{HIV-1 type B in the UK}

We apply the \BDSIR{} method to five HIV-1 clusters sampled between 1999 and 2003, mainly ($85\%$) from men having sex with men (MSM) around London  \citep{Hue:2005fk}.  
Bayesian estimates for the epidemiological parameters and time to the most recent common ancestors of the clusters are summarized in Table \ref{table:hiv_parameters}.

\begin{figure} 
\begin{center}
\includegraphics[width=0.9\linewidth]{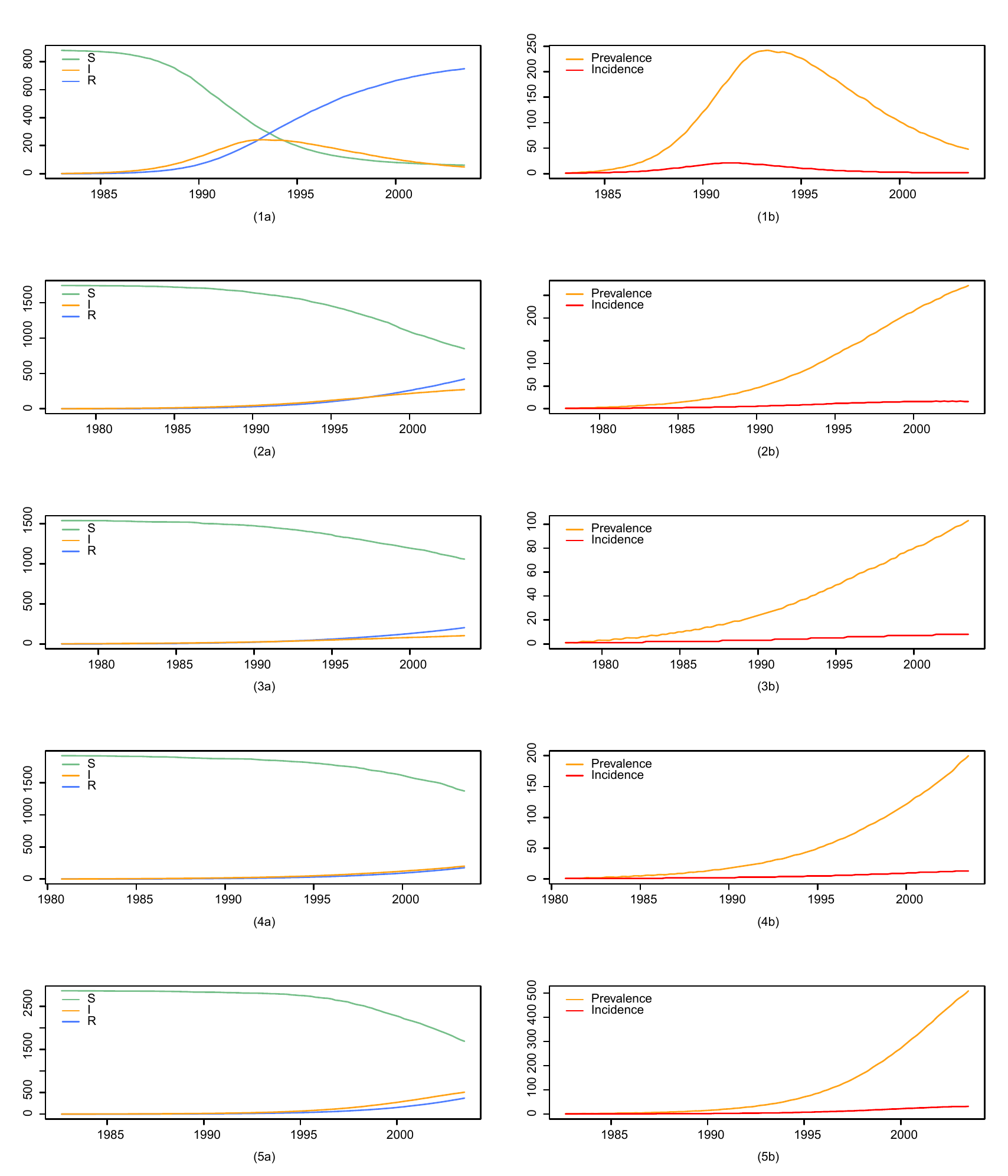}
\caption{{\bf \SIR{} trajectories and incidence of HIV-1 clusters from the United Kingdom} Bayesian posterior mean trajectories for clusters (1-5): The overall \SIR{} dynamics (a) show at what stage in the epidemic each cluster was sampled. Zooming into the number of infecteds, i.e.~the prevalence over time in (b) enables comparison to the incidence.}
\label{fig:bdsir_uk_a}
\end{center}
\end{figure}

\begin{figure} 
\begin{center}
\includegraphics[width=0.6\linewidth]{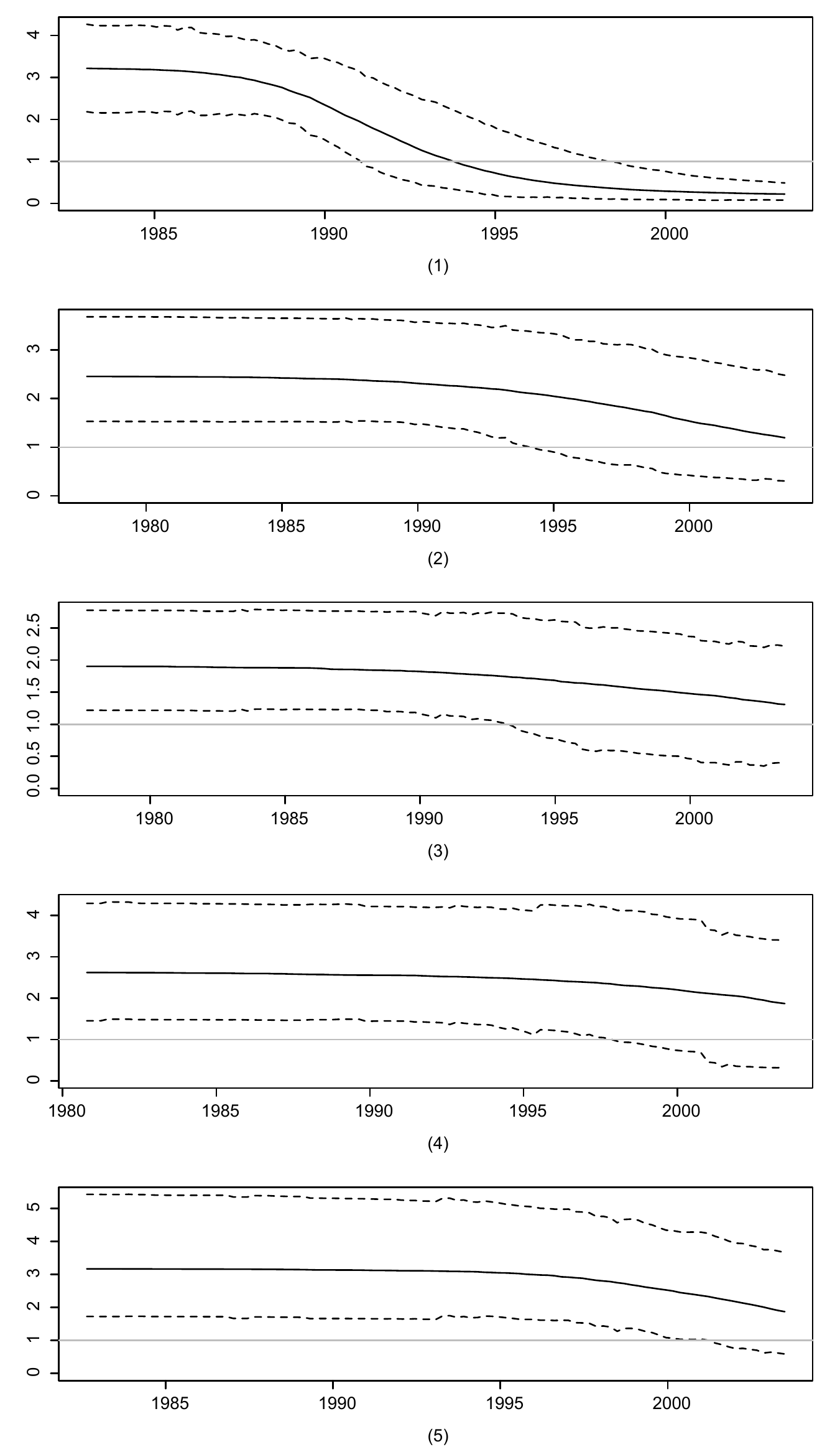}  
\caption{{\bf HIV from the United Kingdom: Reconstructed effective reproduction ratio over time.} Median effective reproduction ratio for each cluster, computed from the posterior birth-death rates and \SIR{} trajectories. Dotted lines show the 95\% HPD interval.}
\label{fig:bdsir_uk_c}
\end{center}
\end{figure}

{Our results suggest that t}he local epidemics corresponding to each of the five genetic clusters have been sampled at varying epidemic stages. 
Figure \ref{fig:bdsir_uk_a} shows the posterior medians of the epidemic time series and suggests that cluster 1 is the only cluster that has gone through the largest part of its local epidemic.
{A single sampled trajectory for each cluster demonstrates the stochastic noise in the epidemics (Figure S1). }
At the end of the sampled interval the pool of susceptible individuals of this cluster has been depleted nearly completely.
On the other hand, the other 4 clusters are just before or at the peak of the local epidemic. 
The estimated depletion of susceptible individuals especially in cluster 2 indicates that those epidemics have progressed fairly far and one would expect a decline in the number of infected individuals soon after the end of the sampled interval. 
These dynamics can also be seen in the plots of the average effective reproduction ratio $\overline{\mathcal{R}}_i$ over time (Figure \ref{fig:bdsir_uk_c}).

The basic reproduction ratio \R{} estimated from these clusters ranges from 1.{90} ($95\%$ HPD: 1.2{2}-2.78) in cluster 3 to 3.{22} ($95\%$ HPD 2.{18}-4.2{7}) in cluster 1. 
{There are significant differences in the estimated \R{} values across the five clusters, despite them all sharing the same prior, which demonstrates that the sequence data contain substantial information about the basic reproduction ratio.}
{These results are robust to a change of the \R{} prior distribution (data not shown).} 
Median estimates of the rate to become non-infectious range from $0.15 - 0.30$, indicating an average infectious period of about $3-7$ years in these clusters. 

In all clusters the estimates of the sampling proportion $s$ and the initial number of susceptible individuals $n_S(0)$ come with broad $95\%$ HPD intervals. 
The median $n_S(0)$ is between {880} and 2900 among the clusters.
Cluster 1 turns out to be the most informative here, with its $95\%$ HPD ranging from ${140-3600}$ (median {880}).
The least informative is cluster 5 ($95\%$ HPD ${180-16900}$, median 2900), which appears to be (a) sampled from the largest epidemic among the 5 clusters and (b) an epidemic for which all samples included in this analysis have been sampled before the epidemic reached its peak.
Hence, one should aim at acquiring samples covering as much of the duration of an epidemic as possible.

\subsection*{HCV type 2c in Argentina}
\begin{figure} 
\begin{center}
\includegraphics[width=\linewidth]{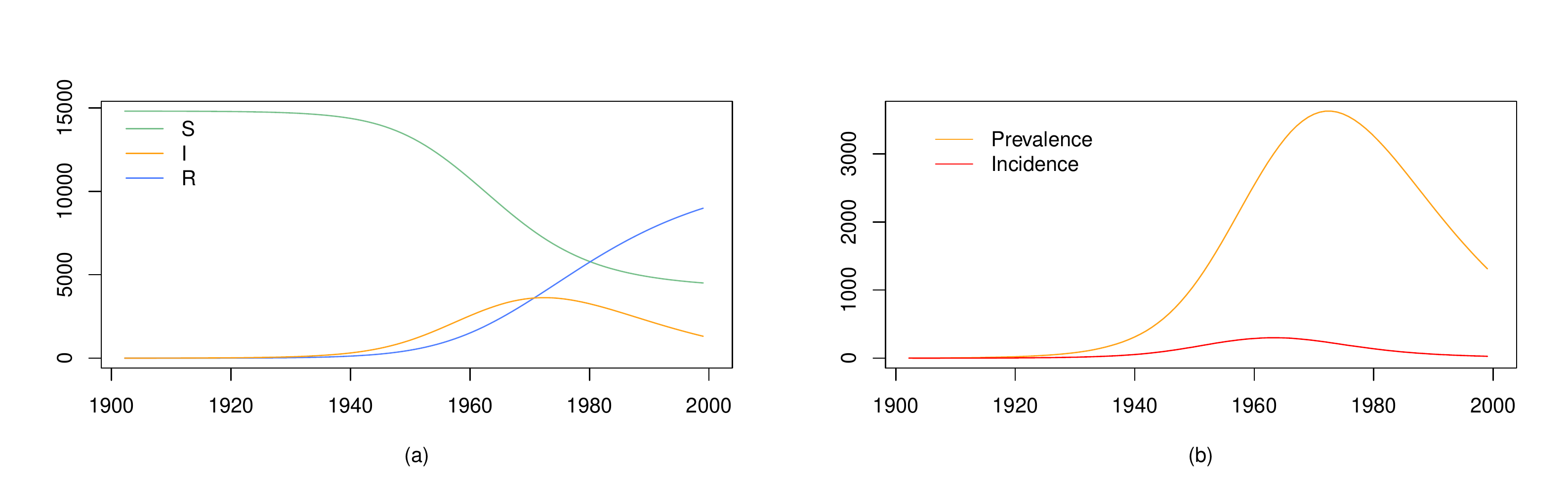}  
\caption{{\bf \SIR{} trajectories and incidence of HCV-2c cluster from CdE, C\'ordoba, Argentina} Bayesian posterior median trajectories: The overall \SIR{} dynamics (a) show that the epidemic peaked around 1970 and is declining since. Zooming into the number of infecteds, i.e.~the prevalence over time in (b) enables comparison to the incidence.}
\label{fig:bdsir_hcv_sir}
\end{center}
\end{figure}

\begin{figure} 
\begin{center}
\includegraphics[width=\linewidth]{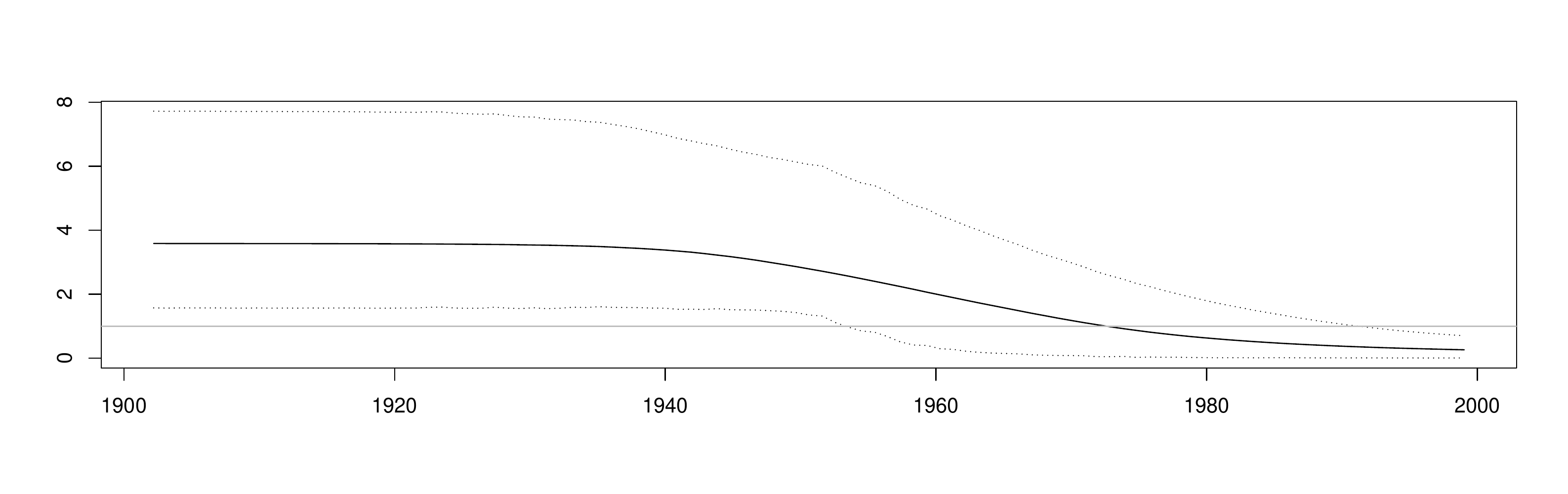}  
\caption{{\bf Reconstructed effective reproduction ratio --- HCV-2c cluster from CdE, C\'ordoba, Argentina} Median effective reproduction ratio, computed from the posterior birth-death rates and \SIR{} trajectories. Dotted lines show the 95 \% HPD interval.}
\label{fig:bdsir_hcv_R}
\end{center}
\end{figure}

{
Applied to a contemporaneously sampled HCV-2c data set from Cruz del Eje (CdE), a city in Argentina, the methods reveals that the virus caused a large local epidemic (Figure \ref{fig:bdsir_hcv_sir}-\ref{fig:bdsir_hcv_R}). 
Despite an uninformative prior distribution on the sampling probability $\rho$, we obtain a median $\rho = 2.6 \%$ (95\% HPD:  2.3\% -  7.6 \%), which agrees very well with direct calculations based on previous estimates \citep{Mengarelli:2006yq}.
We estimate $\mathcal{R}_0 = 3.6 $ (95\% HPD: 1.6 - 7.7),  $n_S(0) = 14800 $ (3200--29600) and $\gamma = 0.056 $ (95\% HPD: 0.014 - 0.134), the latter indicating an infectious period of 17.7 years.
The time of origin of the local epidemic in CdE is estimated to be 1906, with the root of the tree being placed in 1914.
}

{
For the sake of comparability we also analyzed the larger data set (including another 29 sequences from places within C\'ordoba province) that was investigated by \cite{Dearlove:2013ys}. 
Initially, we employed uninformative prior distributions for the epidemiological parameters resulting in an estimate of the epidemic population size of $N=5200$ (400--37000) and a sampling proportion of $s = 68 \%$ (27\%--100\%). 
These results do neither match the large population of C\'ordoba province (1.3 million) nor the small sampling proportion (2.8 \%) encountered by \cite{Mengarelli:2006yq}.
This suggests a model misspecification. 
Given the large size of C\'ordoba province (165 k km$^2$) it appears that this data set requires either the analysis of subsampled local epidemics (as we did for Cruz del Eje) or the incorporation of population structure into the model. 
In fact, repeating the same analysis with a prior distribution that forces the sampling proportion to be small, we obtain results that are very similar to the estimates obtained by \cite{Dearlove:2013ys} under a coalescent \SIR{} model (Figure S3).   
These results might explain why the analysis of the larger set resulted in unrealistically small estimates of the duration for the infectious period (average $1 / \gamma = 1.47$ years (coalescent \SIR), $1 / \gamma = 8.3 $ years (BDSIR)).
}


\section*{Discussion}

Phylodynamic methods play an important role in understanding virus dynamics. 
Awareness of the interaction of evolutionary and ecological dynamics is essential for the development of containment strategies for virus outbreaks over short and long timescales. 
We have presented a model that couples evolutionary processes with the underlying stochastic host dynamics in order to obtain realistic estimates of the evolutionary as well as epidemiological history. 
Existing phylodynamic approaches often infer a phylogeny which is then assumed to be fixed for epidemiological inference \citep{Rasmussen:2011fk, Volz:2012kx}; (see \cite{Kuhnert:2011vn} for a review of further methods).

Our approach couples a birth-death tree prior with a compartmental epidemiological \SIR{} model such that the epidemiological parameters are estimated simultaneously with the reconstruction of the phylogeny. This way the uncertainty of the tree is integrated into the inference of the epidemiological dynamics. 
The choice of the birth-death skyline model as a kernel for the prior on the phylogeny is natural: Epidemiological parameters such as the basic reproduction ratio \R{} are readily computed from an appropriate parametrization, and limitations of the coalescent process, such as the {deterministic population} size assumption, are avoided. 
Note that the assumption of the birth-death skyline plot \citep{Stadler:2013fk}, stating that infected individuals become non-infectious upon sampling, also applies here.  
{This is a somewhat artificial assumption made for computational convenience. To avoid such an assumption would require allowing phylogenetic trees containing ``direct ancestors''}. The first steps towards relaxation of this assumption have {recently} been {taken} \citep{Gavryushkina:2013uq}.

{Recently, \cite{leventhal2013using} developed a similar phylodynamic model that couples a birth--death process with a compartmental \SI{} model, and showed that negligence of the stochastic epidemiological dynamics can introduce bias into phylogenetic reconstruction.}

Traditional coalescent based approaches often suffer from difficulties interpreting the effective population size \citep{Frost:2010fk}.
Explicit simulation of the stochastic \SIR{} trajectories in the \BDSIR{} model yields separate estimates of incidence and prevalence. 
{This explicit separation of incidence and prevalence facilitates correct interpretation of results, although one must still take quantities such as offspring distribution, population structure and selection pressures into account.
Nevertheless}, the resulting trajectories provide information about features such as the time of the epidemic peak.
{Alternatives to the independence MH sampler utilized to sample the stochastic \SIR{} trajectories, like particle filtering \cite{Rasmussen:2011fk} or pure Monte Carlo methods, might yield some computational benefit, but at the expense of the inference of the marginal posterior distribution of the compartment trajectories.}

A promising {coalescent-based} phylodynamic model that 
{incorporates complex population dynamics was developed by \cite{Volz:2012kx}. However, it still assumes a deterministically changing population size.}
In fact, when applied in \citep{Volz:2012fk}, it is based on a fixed phylogeny that has presumably been reconstructed based on a standard coalescent tree prior.
{However, note that \cite{Volz:2012kx} could be extended to take into account stochastic epidemiological dynamics in a similar manner to that employed for the \BDSIR{} model. If stochastic trajectories were used for the coalescent rates and implemented in a Bayesian framework it would enable direct comparison between birth-death methods and the coalescent-based methods described in \cite{Volz:2012kx}. }

In our simulation study we have shown that the \BDSIR{} model accurately estimates epidemiological parameters from simulated \SIR{} trees. 
We have applied the model to five genetic clusters of HIV-1 type B from the United Kingdom. 
The data analysis revealed the epidemic stages in which the clusters were sampled.
Only cluster 1 appears to be at the end of the epidemic, while the other four clusters were sampled around the time of their peak. 
Surprisingly, there is considerable variation in the estimates of the basic reproduction ratio \R{} among the clusters. 
In cluster 3 the estimated median is $1.9$, in cluster 1 and 5 it is slightly above 3. 
{These differences in the estimated \R{} values across the five clusters, and their deviation from the common prior distribution, confirm that the sequence data contain information about the epidemiological parameters.}
Although we did not model variation of the underlying transmission rate among individuals , the variation of estimated epidemiological parameters among the clusters might point us towards the existence of super-spreaders.

Comparing the results of the analysis of cluster 2 to those using the birth-death-skyline plot, published by \cite{Stadler:2013fk}, the estimates of the sampling proportion in both analyses agree ($4{7}\%$ here vs.~$50\%$ BDSKY). 
Expectedly, the estimated basic reproduction ratio $\mathcal{R}_0 = 2.45$ is slightly larger than the effective reproduction ratio $\mathcal{R}_1 = 2.37$ near the origin that resulted from the BDSKY analysis.
Overall, analysis under the parametric \BDSIR{} method resulted in narrower HPD intervals than analysis under the non-parametric BDSKY method, with the \BDSIR{} intervals being contained in the BDSKY intervals.

{
The analysis of 44 HCV-2c sequences from the city of Cruz del Eje supports the theory that this genotype has been introduced to Argentina during a European immigration wave between 1880--1920, since the most recent common ancestor of the sample analyzed here is placed in this period. 
From the CdE subset we have estimated an average duration of infectiousness of 17.7 years, which agrees with the 10-30 year range that has previously been supposed \citep{Pybus:2001fk}.
}

In conclusion, the \BDSIR{} model provides the ability to simultaneously reconstruct evolutionary processes with their underlying host population dynamics from viral sequence data and in particular the inferred parameters allow us to make statements about the future fate of the epidemic.  
{Although we have used strong simplifications concerning the epidemiological dynamics of viruses like HIV (see e.g.~\cite{Eaton:2012fk}), this work is a first step towards more sophisticated methods, and future work shall relax the simplifying assumptions made here.}
We emphasize that this general technique is applicable not only to viruses but to any rapidly evolving organism for which the evolutionary dynamics act on the same timescale as the population processes of their hosts. 
{Future work will aim at extensions that incorporate temporal and spatial structuring of the host and/or viral population.}

\section*{Acknowledgments}
We thank David Welch and Andr\'es Culasso for valuable feedback and discussions and St\'ephane Hu\'e for providing the HIV data sets. 
D.K. and A.J.D. were supported by Marsden Grant UOA0809 and A.J.D. by a Rutherford Discovery Fellowship, both from the Royal Society of New Zealand.
D.K. also thanks ETH Z\"urich, T.S. thanks the Swiss National Science Foundation grant PZ00P3 136820 and ETH Z\"urich, and T.G.V. thanks the Allan Wilson Centre for funding. 
{The authors also wish to acknowledge the contribution of the NeSI high-performance computing facilities and the staff at the Centre for eResearch at the University of Auckland.}

\bibliographystyle{royalb}
\bibliography{BDSIR}

\section*{Tables}

\begin{table}
\begin{tabular}{|c|c|c|c|c|c|c|}
  \hline
& truth & median & error & bias & relative & 95\% HPD  \\ 
&  &  &  &  & HPD width & accuracy (\%) \\ 
  \hline
\R{} & 2.50 & 2.74 & 0.13 & 0.10 & 0.81 & 100.00 \\ 
  $\gamma$ & 0.30 & 0.26 & 0.16 & -0.14 & 0.90 & 99.00 \\ 
  $s$ & 0.17 & 0.22 & 0.34 & 0.32 & 1.90 & 100.00 \\ 
  \hline
\end{tabular}
\caption{{\bf \BDSIR{} simulation results ($n_S(0)$ fixed)}
{Posterior parameter estimates and accuracy obtained from 100 simulated trees with 100 tips sampled sequentially through time. $n_S(0)$ is fixed to the true simulation value. 
For each parameter, the median over the 100 medians / errors / biases / HPD widths / HPD accuracies is provided.} }
\label{table:SIR_BDSIR_fixS}
\end{table}
\begin{table}
\begin{tabular}{|c|c|c|c|c|c|c|}
  \hline
& truth & median & error & bias & relative & 95\% HPD  \\ 
&  &  &  &  & HPD width & accuracy  (\%) \\ 
   \hline
$\overline{\mathcal{R}}_{1}$ & 2.49 & 2.76 & 0.15 & 0.12 & 0.81 & 100.00 \\ 
  $\overline{\mathcal{R}}_{2}$ & 2.48 & 2.73 & 0.15 & 0.12 & 0.81 & 100.00 \\ 
  $\overline{\mathcal{R}}_{3}$ & 2.45 & 2.69 & 0.16 & 0.13 & 0.80 & 100.00 \\ 
  $\overline{\mathcal{R}}_{4}$ & 2.39 & 2.58 & 0.18 & 0.16 & 0.80 & 99.00 \\ 
  $\overline{\mathcal{R}}_{5}$ & 2.25 & 2.42 & 0.24 & 0.24 & 0.79 & 98.70 \\ 
  $\overline{\mathcal{R}}_{6}$ & 2.00 & 2.12 & 0.39 & 0.39 & 0.77 & 97.20 \\ 
  $\overline{\mathcal{R}}_{7}$ & 1.63 & 1.72 & 0.72 & 0.72 & 0.77 & 94.70 \\ 
  $\overline{\mathcal{R}}_{8}$ & 1.23 & 1.32 & 1.27 & 1.27 & 0.77 & 87.50 \\ 
  $\overline{\mathcal{R}}_{9}$ & 0.89 & 0.98 & 2.17 & 2.17 & 0.81 & 83.90 \\ 
  $\overline{\mathcal{R}}_{10}$ & 0.65 & 0.76 & 3.33 & 3.33 & 0.86 & 80.67 \\ 
   \hline
\end{tabular}
\caption{{\bf \BDSIR{} simulation results ($n_S(0)$ fixed)}
{Computed averages for the effective reproduction number 
from 100 simulated trees with 100 tips sampled sequentially through time. $n_S(0)$ is fixed to the true simulation value. 
For each parameter, the median over the 100 medians / errors / biases / HPD widths / HPD accuracies is provided. 
The averages $\mathcal{\overline{R}}_i$ for $i=1..10$ were computed from the estimated trajectories, \R{},  $\gamma$ and $s$.} }
\label{table:SIR_BDSIR_fixS_averages}
\end{table}
\begin{table}
\begin{tabular}{|c|c|c|c|c|c|c|}
  \hline
& truth & median & error & bias & relative & 95\% HPD  \\ 
&  &  &  &  & HPD width & accuracy  (\%) \\ 
  \hline
\R{} & 2.50 & 2.63 & 0.12 & 0.05 & 0.87 & 100.00 \\ 
  $\gamma$ & 0.30 & 0.29 & 0.13 & -0.05 & 1.21 & 100.00 \\ 
  $s$ & 0.17 & 0.18 & 0.19 & 0.11 & 1.95 & 100.00 \\ 
  $n_S(0)$ & 999.00 & 1900.68 & 0.90 & 0.90 & 5.44 & 100.00 \\ 
  \hline
\end{tabular}
\caption{{\bf \BDSIR{} simulation results ($n_S(0)$ estimated)}
{Posterior parameter estimates and accuracy obtained from 100 simulated trees with 100 tips sampled sequentially through time. $n_S(0)$ is estimated in each analysis. 
For each parameter, the median over the 100 medians / errors / biases / HPD widths / HPD accuracies is provided.} }
\label{table:SIR_BDSIR_estS}
\end{table}
\begin{table}
\begin{tabular}{|c|c|c|c|c|c|c|}
  \hline
& truth & median & error & bias & relative & 95\% HPD  \\ 
&  &  &  &  & HPD width & accuracy  (\%) \\ 
   \hline
$\overline{\mathcal{R}}_{1}$ & 2.49 & 2.64 & 0.13 & 0.07 & 3.10 & 100.00 \\ 
  $\overline{\mathcal{R}}_{2}$ & 2.48 & 2.62 & 0.13 & 0.08 & 3.10 & 100.00 \\ 
  $\overline{\mathcal{R}}_{3}$ & 2.45 & 2.58 & 0.14 & 0.09 & 3.10 & 100.00 \\ 
  $\overline{\mathcal{R}}_{4}$ & 2.39 & 2.47 & 0.15 & 0.12 & 3.08 & 100.00 \\ 
  $\overline{\mathcal{R}}_{5}$ & 2.25 & 2.33 & 0.20 & 0.19 & 3.07 & 100.00 \\ 
  $\overline{\mathcal{R}}_{6}$ & 2.00 & 2.07 & 0.34 & 0.34 & 3.04 & 100.00 \\ 
  $\overline{\mathcal{R}}_{7}$ & 1.63 & 1.70 & 0.65 & 0.65 & 3.06 & 100.00 \\ 
  $\overline{\mathcal{R}}_{8}$ & 1.23 & 1.31 & 1.19 & 1.19 & 3.11 & 100.00 \\ 
  $\overline{\mathcal{R}}_{9}$ & 0.89 & 0.97 & 2.05 & 2.05 & 3.26 & 100.00 \\ 
  $\overline{\mathcal{R}}_{10}$ & 0.65 & 0.75 & 3.16 & 3.16 & 3.47 & 100.00 \\ 
   \hline
\end{tabular}
\caption{{\bf \BDSIR{} simulation results ($n_S(0)$ estimated)}
{Computed averages for the effective reproduction number
 from 100 simulated trees with 100 tips sampled sequentially through time. $n_S(0)$ is estimated in each analysis. For each parameter, the median over the 100 medians / errors / biases / HPD widths / HPD accuracies is provided. The averages $\mathcal{\overline{R}}_i$ for $i=1..10$ were computed from the estimated trajectories, \R{},  $\gamma$ and $s$.} }
\label{table:SIR_BDSIR_estS_averages}
\end{table}

\begin{table}
\begin{tabular}{|c|c|c|c|c|c|c|}
  \hline
& truth & median & error & bias & relative & 95\% HPD  \\ 
&  &  &  &  & HPD width & accuracy  (\%) \\ 
  \hline
  $\gamma$ & 0.30 & 0.23 & 0.24 & -0.23 & 0.28 & 99\\ 
 $s$ &  $0.1\overline{6}$ & 0.24 & 0.46 & 0.44 & 0.40 & 100 \\ 
  \hline
$\mathcal{\overline{R}}_1$ & 2.49 & 2.49 & 0.33 & -0.003 & 5.81 & 100\\ 
  $\mathcal{\overline{R}}_2$ & 2.48 & 2.53 & 0.32 & 0.02 & 4.88 & 99\\ 
  $\mathcal{\overline{R}}_3$ & 2.45 & 2.72 & 0.30 & 0.11 & 4.13 & 99\\ 
  $\mathcal{\overline{R}}_4$ & 2.39 & 2.73 & 0.27 & 0.14 & 3.33 & 98\\ 
  $\mathcal{\overline{R}}_5$ & 2.25 & 2.65 & 0.24 & 0.17 & 2.77 & 97\\ 
  $\mathcal{\overline{R}}_6$ & 2.00 & 2.31 & 0.23 & 0.14 & 2.24 & 95\\ 
  $\mathcal{\overline{R}}_7$ & 1.63 & 1.85 & 0.27 & 0.11 & 1.90 & 92\\ 
  $\mathcal{\overline{R}}_8$ & 1.23 & 1.42 & 0.32 & 0.12 & 1.69 & 91\\ 
  $\mathcal{\overline{R}}_9$ & 0.89 & 1.01 & 0.32 & 0.11 & 1.58 & 98\\ 
  $\mathcal{\overline{R}}_{10}$ & 0.65 & 1.16 & 0.77 & 0.77 & 2.26 & 97\\ 
   \hline
\end{tabular}
\caption{{\bf Birth-death skyline simulation results}
{Birth-death skyline posterior parameter estimates and accuracy obtained from 100 simulated trees with 100 tips sampled sequentially through time. Rate changes are allowed among 10 equidistant intervals. For each parameter, the median over the 100 medians / errors / biases / HPD widths / HPD accuracies is provided.} }
\label{table:SIR_BDSKY_10}
\end{table}

\begin{table}
\begin{tabular}{|c|c|c|c|c|c|c|}
  \hline
& truth & median & error & bias & relative & 95\% HPD  \\ 
&  &  &  &  & HPD width & accuracy  (\%) \\ 
  \hline
$\mathcal{\overline{R}}$ & 1.86 & 1.63 & 0.13 & -0.12 & 1.17 & 92 \\ 
  $\gamma$ & 0.30 & 0.30 & 0.08 & -0.002 & 0.52 & 100 \\ 
  $s$ & $0.1\overline{6}$ & 0.17 & 0.04 & 0.02 & 0.38 & 100 \\ 
 \hline
\end{tabular}
\caption{{\bf Birth-death-sampling simulation results}
{Birth-death-sampling posterior parameter estimates and accuracy obtained from 100 simulated trees with 100 tips sampled sequentially through time. Rates are assumed constant over time. For each parameter, the median over the 100 medians / errors / biases / HPD widths / HPD accuracies is provided.}}
\label{table:SIR_BDSKY_1}
\end{table}

\begin{table}
\begin{tabular}{|  c | c|c|c|c|c| c|c|}
\hline
 Cluster & \R{} & $\gamma$ & $s$ & $n_S(0)$ & Root of the & Origin of the \\
  &&  &  &  & tree (yr) &  epidemic (yr)\\
  \hline
 1 & 3.22 & 0.30 & 0.68 & 880 & 1986 & 1983 \\ 
   & (2.18-4.27) & (0.15-0.47) & (0.25-1) & (142-3592) & (1983-1988) & (1978-1987) \\ 
  \hline
  2 & 2.45 & 0.17 & 0.47 & 1745 & 1983 & 1978 \\ 
   & (1.53-3.68) & (0.06-0.35) & (0.1-0.96) & (190-8892) & (1979-1986) & (1968-1984) \\ 
  \hline
  3 & 1.90 & 0.20 & 0.68 & 1540 & 1985 & 1978 \\ 
   & (1.22-2.78) & (0.09-0.39) & (0.27-1) & (153-8558) & (1981-1988) & (1962-1986) \\ 
  \hline
  4 & 2.62 & 0.15 & 0.38 & 1921 & 1987 & 1981 \\ 
   & (1.45-4.29) & (0.06-0.31) & (0.06-0.93) & (128-11007) & (1983-1990) & (1970-1988) \\ 
  \hline
  5 & 3.17 & 0.15 & 0.21 & 2862 & 1986 & 1983 \\ 
   & (1.73-5.43) & (0.06-0.31) & (0.02-0.79) & (183-16909) & (1981-1989) & (1975-1989) \\ 
\hline
\end{tabular}
\caption{{\bf {HIV-1 type B from the UK: Bayesian parameter estimates}}
{Bayesian parameter estimates and HPD intervals (in parentheses) from phylodynamic analysis of five HIV-1 type B cluster from the United Kingdom.} }
\label{table:hiv_parameters}
\end{table}

\begin{table}
\begin{tabular}{|  c | c|c|c|c|c| c|}
\hline
 Analysis & \R{} & $\gamma$ & $s$ & $n_S(0)$ & {$T$} & $\rho$\\
  \hline
Simulated \SIR{} & LogN(1,1) & LogN(-0.5,1) & Beta(2,10) & LogN(7,1) & - & - \\
HIV data UK & LogN(0.5,0.5) & LogN(-1,0.75) & Beta(1,1) & LogN(7,1.25) & Unif(0,{1000}) & -\\ 
HCV data CdE & LogN(0,2) & LogN(-0.5,1.25) & - & Unif(0,{30000}) & Unif(0,{1000}) & Unif(0,1)\\ 
   \hline
\end{tabular}
\caption{{\bf Prior distributions}
{Prior distributions for the re-estimation of \SIR{} parameters from simulated trees (equal priors applied in \BDSIR{} and birth-death-sampling analyses), and for data analyses.} }
\label{table:priors}
\end{table}

\endthebibliography

\end{document}